\documentclass[conference]{IEEEtran}
\IEEEoverridecommandlockouts
\usepackage{cite}
\usepackage{amsmath,amssymb,amsfonts}
\usepackage{algorithmic}
\usepackage{graphicx}
\usepackage{textcomp}
\usepackage{xcolor}
\usepackage{xcolor}
\usepackage{booktabs}
\usepackage[skip=1pt]{caption}
\usepackage{enumitem}
\setlist[itemize]{topsep=2pt, partopsep=0pt, parsep=0pt, itemsep=2pt}
\usepackage{graphicx} 
\usepackage{float} 
\usepackage{titlesec}
\usepackage{stfloats}
\usepackage{booktabs}
\usepackage{times}
\usepackage{flushend}
\usepackage[T1]{fontenc}
\def\BibTeX{{\rm B\kern-.05em{\sc i\kern-.025em b}\kern-.08em
    T\kern-.1667em\lower.7ex\hbox{E}\kern-.125emX}}
\begin{document}

\title{REACH: Reinforcement Learning for Adaptive Microservice Rescheduling in the Cloud–Edge Continuum\\
}

\author{
\IEEEauthorblockN{Xu Bai\IEEEauthorrefmark{1},
Muhammed Tawfiqul Islam\IEEEauthorrefmark{2},
Rajkumar Buyya\IEEEauthorrefmark{3},
Adel N. Toosi\IEEEauthorrefmark{4}}
\IEEEauthorblockA{\IEEEauthorrefmark{1}\IEEEauthorrefmark{2}\IEEEauthorrefmark{3}\IEEEauthorrefmark{4} School of Computing and Information Systems, University of Melbourne, Australia}
\IEEEauthorblockA{\IEEEauthorrefmark{1}bxb1@student.unimelb.edu.au,
\IEEEauthorrefmark{2}tawfiqul.islam@unimelb.edu.au,
\IEEEauthorrefmark{3}rbuyya@unimelb.edu.au,
\IEEEauthorrefmark{4}adel.toosi@unimelb.edu.au}
}


\maketitle
\thispagestyle{plain}
\pagestyle{plain}
\begin{abstract}
Cloud computing, despite its advantages in scalability, may not always fully satisfy the low-latency demands of emerging latency-sensitive pervasive applications. The cloud-edge continuum addresses this by integrating the responsiveness of edge resources with cloud scalability. Microservice Architecture (MSA)  characterized by modular, loosely coupled services, aligns effectively with this continuum. However, the heterogeneous and dynamic computing resource poses significant challenges to the optimal placement of microservices. We propose REACH, a novel rescheduling algorithm that dynamically adapts microservice placement in real time using reinforcement learning to react to fluctuating resource availability, and performance variations across distributed infrastructures. Extensive experiments on a real-world testbed demonstrate that REACH reduces average end-to-end latency by 7.9\%, 10\%, and 8\% across three benchmark MSA applications, while effectively mitigating latency fluctuations and spikes.
\end{abstract}


\section{Introduction}

Cloud computing has transformed information technology with scalable service models, abundant resources, and cost advantages over traditional infrastructures~\cite{gong2010characteristics}. However, the proliferation of IoT devices and AI-driven applications has created pervasive computing scenarios where users demand seamless, low-latency services that centralized cloud data centers alone cannot provide~\cite{shi2016edge}.

To address this limitation, the cloud–edge continuum integrates cloud and edge resources into a unified infrastructure~\cite{moreschini2022cloud}. It combines the scalability of cloud platforms with the proximity of edge devices, enabling applications to flexibly allocate resources across layers and deliver ubiquitous, adaptive, and responsive services. Meanwhile, the widespread adoption of microservice architecture (MSA) has further decomposed applications into fine-grained, loosely coupled services, which are well suited to dynamic deployment across heterogeneous cloud–edge resources.

\begin{figure}[H]
    \centering
    \includegraphics[width=0.9\linewidth]{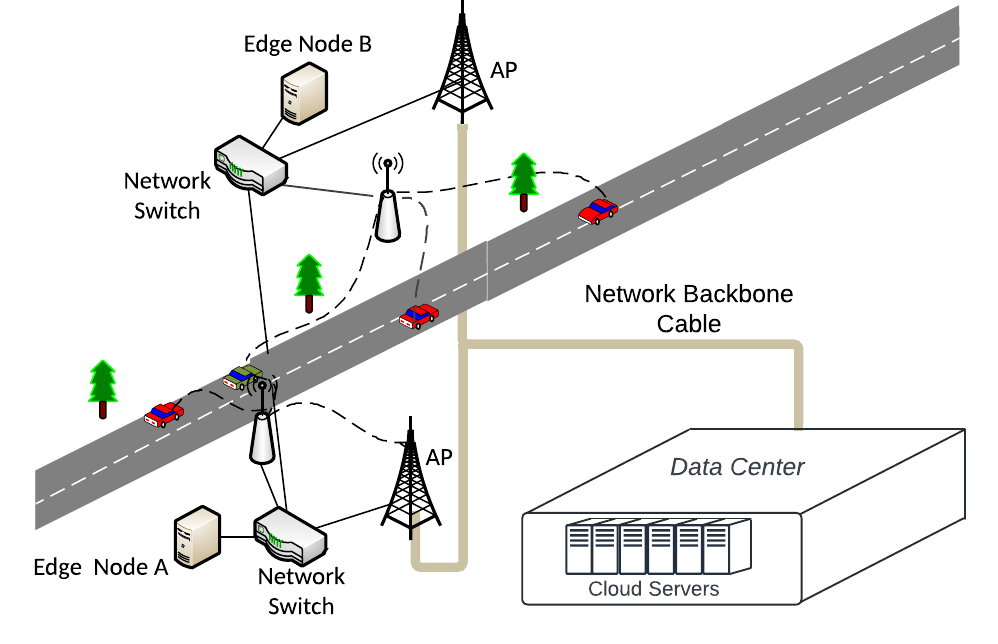}
    \caption{Vehicular computing leveraging the cloud–edge continuum}
    \label{fig:vec-computing}
    \vspace{-0.5cm}
\end{figure}

Figure~\ref{fig:vec-computing} illustrates a vehicular computing scenario within the cloud–edge continuum. Vehicles equipped with on-board units (OBUs) connect to access points (APs), which are linked to edge nodes and data centers through network backbone cables. This architecture allows moving vehicles to exploit both the low-latency response of nearby edge nodes and the high computational capacity of cloud servers for real-time distributed computation. Therefore, adaptively scheduling microservices across heterogeneous resources in the continuum is essential to ensuring both service quality and system safety.

Existing research has explored optimizing application performance in the cloud-edge continuum through various service scheduling strategies, yet several critical challenges remain. First, in a cloud-edge continuum, resources are in constant flux: services scale dynamically, new edge devices join the network, and cloud nodes are added or removed based on current resource utilization. In such an environment, scheduling algorithms need to change scheduling plans frequently to adapt to changes.
Second, microservice scheduling is challenging because of complex service dependencies and diverse characteristics, especially when optimizing placement in the cloud–edge continuum. Reinforcement learning (RL) has been widely applied to this problem but often faces the slow-start issue~\cite{botvinickReinforcementLearningFast2019}, since online training requires real-time execution of each action. Simulation can reduce this cost, but the alignment between policies learned from simulation and the real world is questionable. Thus, an RL pipeline that aligns simulation with the real cloud–edge continuum is crucial for practical deployment.

Finally, evaluating MSA placement strategies in real-world testbeds remains a significant challenge, particularly due to the complexity of deploying heterogeneous computing resources over cloud-edge continuum environments. As a result, most existing studies rely on simulation environments to validate their algorithms. However, such simulations may not fully capture the complexities and variability of real-world deployments, thereby limiting the applicability and effectiveness of these algorithms in production cloud-edge continuum scenarios.

To address these challenges, we propose REACH, an RL-based rescheduling algorithm designed to optimize the end-to-end latency of MSA applications within cloud-edge continuum. REACH models heterogeneous computing resources across the cloud-edge continuum and accounts for the network latency between microservices and computing nodes. Furthermore, it continuously monitors the state of the cloud-edge environment and incrementally reschedules only essential microservices, using minimal steps to reduce overhead, minimize application downtime, and prevent unnecessary disruptions.

Our \textbf{main contribution} are as follows:
\begin{itemize}
    \item We propose a novel MSA rescheduling algorithm that iteratively adjusts microservice placements in response to environmental changes. Unlike most existing approaches, which generate a new placement plan for all microservices without considering rescheduling costs, our method incrementally selects the most probable next rescheduling action for currently running microservices to optimize end-to-end application latency. Moreover, the rescheduling cost is explicitly integrated into our model to ensure that the number of rescheduling actions is minimized.
    \item We designed and implemented a custom RL simulation environment \textbf{CEEnv} that models heterogeneous computing resources and network conditions, reflecting real-world cloud-edge continuum scenarios. 
    \item Rather than relying solely on simulation environments, we implement a deployment pipeline that transfers the strategies learned in simulation directly to a real-world Kubernetes-based cloud-edge continuum testbed. Experimental results demonstrate that, in this real-world setting, our algorithm significantly reduces end-to-end latency, stabilizes response times, and mitigates latency spikes during node failures and sudden traffic surges.
\end{itemize}

\section{Related Work}
To address the Microservice Architecture (MSA) scheduling problem, numerous studies have used heuristic algorithms that consider factors such as computing resource availability, application latency, and network bandwidth usage to generate placement plans \cite{filipMicroservicesSchedulingModel2018a,LightweightDecentralizedService,centofantiLatencyAwareKubernetesScheduling2023}. These heuristics aim to optimize service execution and reduce latency by scheduling microservices closer to end users. However, they often simplify invocation patterns, assuming linear chains or focusing on individual service latency, which fails to capture the complexities of real-world MSA applications.

Optimization-based methods, using techniques such as Mixed Integer Linear Programming (MILP) and Particle Swarm Optimization (PSO), have also been proposed to tackle microservice placement with multiple objectives such as minimizing cost and latency \cite{OptimalDeploymentFog,xieNovelDirectionalNonlocalconvergent2019,alelyaniOptimizingCloudPerformance2024}. These methods formalize the problem mathematically and aim to optimize placement decisions dynamically. Nevertheless, optimization approach often incurs relatively high computational overhead to calculate the optimal placement for every decision making process, and many of them are lack of real-world deployment evaluations, raising concerns about their effectiveness in practical, dynamic cloud-edge continuum. 

Given the complexity and dynamics of MSA placement, reinforcement learning (RL) has been widely explored. Mampage~et~al.~\cite{mampageDeepReinforcementLearning2023a} proposed a deep Q-Network (DQN) model for container scheduling in serverless environments, but it only considers individual containers and ignores interdependencies. Ma~et~al.~\cite{maDeepMultiagentReinforcement2025a} applied a multi-agent RL approach for microservice migration in cloud platforms, showing promising results on synthetic and trace datasets. However, their method is confined to cloud settings and overlooks device heterogeneity which is common in cloud-edge continuum

Maia and Ghamri-Doudane~\cite{maiaDeepReinforcementLearning2023} propose an RL-based method for scalable microservice scheduling across the cloud-edge continuum. However, their model overlooks the heterogeneity of computing resources, which limits its applicability to real-world deployments. Similarly, Chen et al.\cite{chenIoTMicroserviceDeployment2021} introduce MB\_DDPG, an RL algorithm aimed at reducing average request waiting time through adaptive microservice placement. Yet, their evaluation is confined to simulations and assumes a linear invocation pattern, making it unsuitable for complex real-world MSA structures.

Lv et al.\cite{lvGraphReinforcementLearningBasedDependencyAwareMicroservice2024} incorporate Graph Convolutional Networks (GCNs) into a DQN-based scheduler to capture microservice call dependencies, but also neglect resource heterogeneity, which is crucial in cloud-edge environments. Afachao et al.~\cite{afachaoEfficientMicroserviceDeployment2024} propose BAMPP, a variant of the actor-critic framework with improved training stability, but their evaluation is similarly limited to only simulated environments.

In summary, existing heuristic-based approaches rely on simplified criteria such as latency or user proximity, often overlooking complex invocation patterns and resource heterogeneity, which leads to suboptimal microservice placements. While optimization methods can yield near-optimal results, they require detailed modeling and are typically too computationally intensive for practical deployment. Existing RL approaches generally generate a one-time scheduling plan for all microservices, neglecting the overhead associated with redeploying services in dynamic cloud-edge environments—such as during node failures, workload surges, or network fluctuations. Additionally, most RL-based studies are confined to simulation environments.
\begin{table}[H]
\caption{Comparison of the Related Works}
\label{table:cmp}
\centering
\resizebox{\columnwidth}{!}{%
\begin{tabular}{@{}ccccc@{}}
\toprule
Work & \begin{tabular}[c]{@{}c@{}}Overhead-Aware \\ Rescheduling\end{tabular} & \begin{tabular}[c]{@{}c@{}}Real-World \\ System\end{tabular} & \begin{tabular}[c]{@{}c@{}}Cloud-Edge\\ Continuum Compliant\end{tabular} & \begin{tabular}[c]{@{}c@{}}Placement \\ Algorithm\end{tabular} \\ \midrule
\cite{fuAdaptiveResourceEfficient2022} &  & \checkmark & \checkmark & RL \\
\cite{pallewattaQoSawarePlacementMicroservicesbased2022} &  &  &  & Meta-Heuristic \\
\cite{faticantiDeploymentApplicationMicroservices2020} &  &  & \checkmark & Heuristic \\
\cite{lvGraphReinforcementLearningBasedDependencyAwareMicroservice2024} &  &  &  & RL \\
\cite{armaniCostEffectiveWorkloadAllocation2021} &  &  & \checkmark & Heuristic \\
\cite{xieNovelDirectionalNonlocalconvergent2019} &  &  & \checkmark & Meta-Heuristic \\
\cite{centofantiLatencyAwareKubernetesScheduling2023} &  & \checkmark &  & Heuristic \\
\cite{chenIoTMicroserviceDeployment2021} &  &  & \checkmark & RL \\
\cite{afachaoEfficientMicroserviceDeployment2024} &  &  & \checkmark & RL \\
\textbf{This work} & \checkmark & \checkmark & \checkmark & RL \\ \bottomrule
\end{tabular}%
}
\end{table}

In addition to addressing the aforementioned gaps, our work is, to the best of our knowledge, the first to demonstrate the practical effectiveness of a learned RL-based microservice rescheduling policy in a real-world cloud-edge continuum testbed with heterogeneous computing nodes. A comparative summary with related studies is provided in Table~\ref{table:cmp}.

\section{Microservice Application Scheduling in Cloud-Edge Continuum}
\label{title:problemformation}


\subsection{\textbf{System Model}}
\label{title:system-model}

Scheduling a microservice on the cloud-edge continuum directly affects its execution time and latency to end users. Cloud nodes, with faster CPUs and more memory, shorten execution time, while edge nodes reduce network latency due to proximity. Furthermore, the position of the microservice relative to its calling and callee services shapes the network topology, affecting the overall end-to-end latency. 

\begin{figure}[H]
    \centering
    \includegraphics[width=0.65\linewidth]{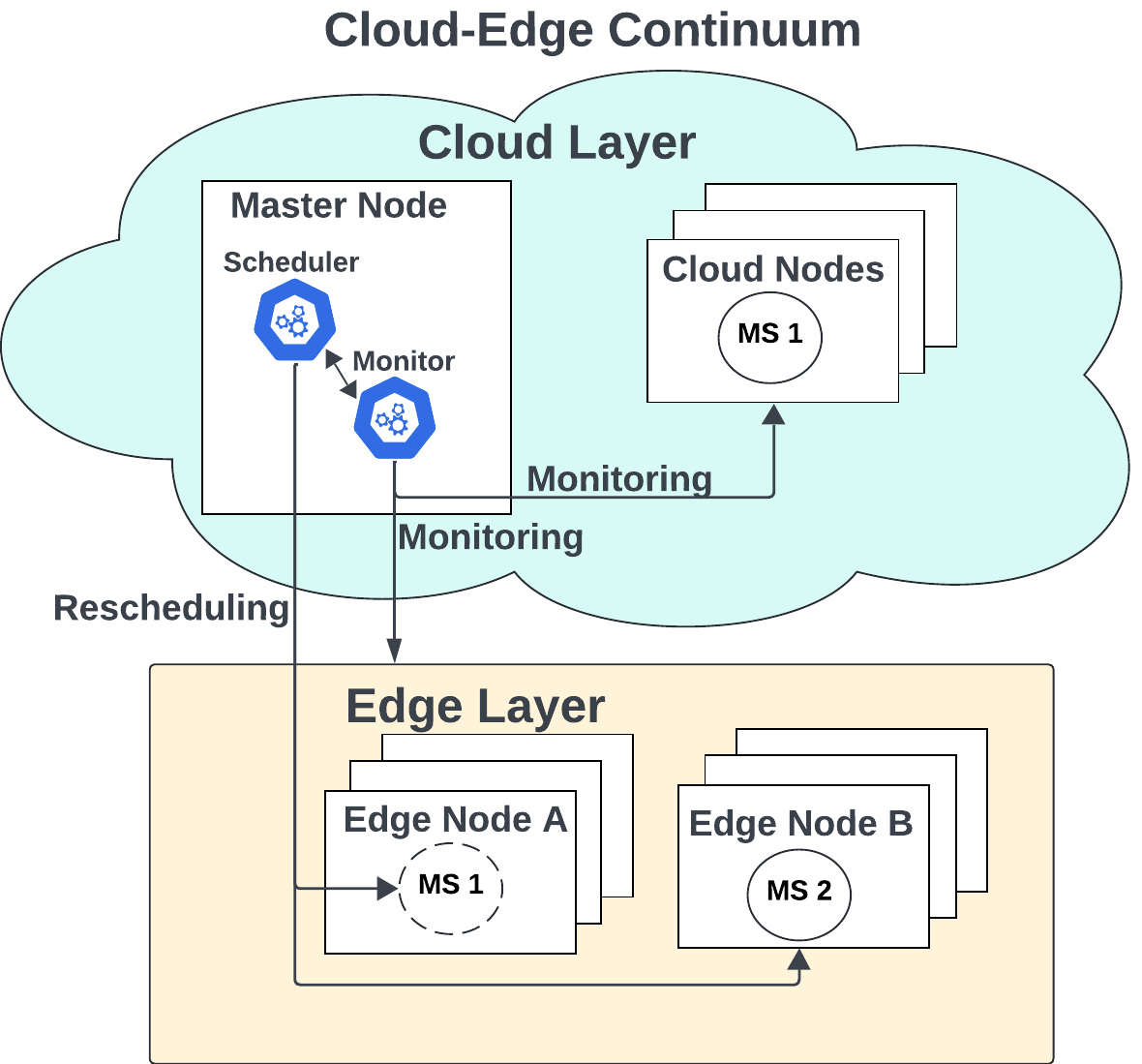}
    \caption{System Architecture}
    \label{fig:problem-arch}
\end{figure}

Figure~\ref{fig:problem-arch} presents the architecture of the proposed system model, where cloud and edge resources are collectively managed by a container orchestrator, forming a unified pool of heterogeneous computing resources. At the core of this architecture lies the scheduler component, which manages the placement of microservices. 

Initially, the scheduler deploys each service within the MSA application according to a predefined scheduling policy. Then scheduler continuously monitors both the state of the system and the operational status of the MSA applications and chooses to progressively reschedule microservices that utilize RL to minimize end-to-end latency and enhance overall performance of the MSA applications. This ensures that the system can adapt effectively to dynamic conditions, maintaining optimal placement as resource availability fluctuates.

To accurately represent real-world deployments of MSA applications within the cloud-edge continuum, we model each service as potentially having multiple replicas to facilitate autoscaling and ensure fault tolerance. Each replicated instance of a service is referred to as a \textbf{Service Instance,} and a collection of these instances constitutes a \textbf{Service}.

In the proposed system model, the cloud-edge continuum consists of heterogeneous computing nodes. Each node \( i \) has CPU and memory capacities denoted as \( C_{\text{cpu}}^i \) and \( C_{\text{mem}}^i \). We assume that the total CPU and memory resources requested by all service instances on any node do not exceed its capacity. This is typically enforced by container orchestration platforms to prevent resource contention. Thus, for any node \( i \), the requested CPU and memory resources, \( R_{\text{cpu}}^i \) and \( R_{\text{mem}}^i \), must satisfy the following constraints:

\begin{equation}
\sum_{i=1}^{n}  R_{{cpu}}^i \leq C_{{cpu}}^i \quad {,} \quad \sum_{i=1}^{n} R_{{mem}}^i \leq C_{{mem}}^i
\end{equation}


To account for computational heterogeneity across the continuum, we define the average execution time \(E_{n_i}^{i}\) for a service instance \(i\) at node \(n_i\), which reflects variations in CPU speed, disk I/O, and memory performance. The heterogeneous network condition is modeled by the inter-node latency \(D_{n_i n_j}\) within the cloud--edge continuum. Accordingly, service-to-service latency is modeled based on the latency of the nodes on which the services are deployed.




\begin{table}[t]
\caption{Notation Table}
\label{table:notion}
\resizebox{\columnwidth}{!}{%
\begin{tabular}{@{}cl@{}}
\toprule
\textbf{Symbol}   & \textbf{Definition}                                                                                                                        \\ \midrule
\(\beta\)         & The index set of all services                                                                                                               \\
\(\tau\)          & The index set of all computing nodes                                                                                                        \\
$\psi_{i} $         & The index set of service instances, \(i\in\beta\)                                                                                    \\
 \(N\)&Number of Nodes in cloud-edge continuum\\
$g_{i}$        & The index set of service groups, \(i\in\beta\)                                                                                              \\
\(E_{i}\)         & Set of all external services called by a service, \(i\in\beta\)                                                                             \\
\( S_{{cur}}^i \) & Current number of running instances in service \(i\in\beta\)                                                                                \\
\( S_{{max}}\)    & Maximum number of running instances in MSA application.                                                                                     \\
\( S_{cur}\)      & Set of all service instances in the running states                                                                                          \\
\( C_{{cpu}}^i \) & CPU capacity of a node \(i\in\tau\)                                                                                                         \\
\( C_{{mem}}^i \) & Memory capacity of a node \(i\in\tau\)                                                                                                      \\
\( A_{{cpu}}^i \) & CPU availability of a node \(i\in\tau\)                                                                                                     \\
\( A_{{mem}}^i \) & Memory availability of a node \(i\in\tau\)                                                                                                  \\
\( R_{{cpu}}^i \) & Total Requested CPU of a node \(i\in\tau\)                                                                                                  \\
\( R_{{mem}}^i \) & Total Requested Memory of a node i \(i\in\tau\)                                                                                             \\
\(D_{i,j}\)       & Latency between two nodes \(i\in\tau\), \(j\in\tau\)                                                                                        \\
\(D_{user}^{i}\)  & Latency between end user and a nodes \(i\in\tau\)                                                                                           \\
\( D_{{msa}}\)    & End-to-end MSA application latency                                                                                                          \\
\(D_{{S}}^{ij} \) & Average latency between services \( i \) and \( j \). \(i\in\beta\), \(j\in\beta\)                                                          \\
$D_{{gateway}}$   & The average latency between end users and the gateway service.   \\ 
\(t_i \)          & Internal execution time of running service instance \(i \in \psi\)                                                                          \\
\(E_{j}^{i}\)     & \begin{tabular}[c]{@{}l@{}}Internal execution time of service instance i running in node j. \\ \(i \in \psi\), \(j \in \tau \)\end{tabular} \\
\( d_{ij}\)       & Latency between two service instances \(i \in \psi\)                                                                                        \\
\(n_i\)           & Deployed node for a service instance \(i \in \psi\)                                                                                         \\
\( r_{{cpu}}^i \) & Requested cpu resource of service instance \(i \in \psi\)                                                                                   \\
\( r_{{mem}}^i \) & Requested memory resource of service instance \(i \in \psi\)                                                                                \\
$T_{ij}$ &        Average invocation time between service $i$ and $j$. \(i\in\beta\), \(j\in\beta\)  \\
\(T_{{E}}^{i}\)   & Average internal execution time of service \(i\),                                                                                           \\
\(T_{{G}}^{i} \)  & Total Service Invocation Time of Service Group \(i \in \omega\)                                                                             \\
\( T_{{S}}^i \)   & Average service processing time for service \(i\in\beta\)                                                                                   \\ \bottomrule
\end{tabular}%
}
\vspace{-0.3cm}
\end{table}

\subsection{\textbf{Problem Formulation}}
\label{title:problem-formulation}
We formulate our problem as rescheduling service instances to minimize end-to-end user request latency. The notations used in this document are summarized in Table~\ref{table:notion}. The end-to-end latency comprises the accumulated execution time of services invoked in processing a single-user request and the network latency incurred when the request is transmitted between services. The internal execution time of a service \(i\) on its deployed node \(n_i\) is denoted by \(E_{{n}_i}^{i}\). For simplicity, we define \(t_i = E_{{n}_i}^{i}\). Then the average internal execution time of service \(i\), represented as \( T_{{E}}^{i} \) is formulated as:
\begin{equation}
\label{eq:exec}
T_{{E}}^{i} = \frac{1}{S_{{cur}}^i} \sum_{j \in \psi_i} {t}_{j} 
\end{equation}
where \( S_{{cur}}^i \) denotes the total number of current running service instances in service \( i \).

To estimate the expected latency between two services, we calculate the average latency across all instances that interact between them. The latency between instances \( k \) and \( l \) is defined as \( d_{kl} = D_{n_k n_l} \), where \( D_{n_i n_j} \) denotes the node-to-node latency between nodes \( n_i \) and \( n_j \). Thus, for services \( i \) and \( j \), the total latency across all instance pairs is:


\begin{equation}
D_{\text{sum}}^{ij} = \sum_{k \in \psi_i} \sum_{l \in \psi_j} d_{kl}
\end{equation}

By dividing $D_{\text{sum}}^{ij}$ by the total number of pairs of service instances - which is the product of the number of service instances \( S_{cur}^i \) and \( S_{cur}^j \) in each service - we obtain the average latency between services \( i \) and \( j \):

\begin{equation}
{D}_{{S}}^{ij} = \frac{1}{{S}_{{cur}}^i \cdot {S}_{{cur}}^j} {D}_{{sum}}^{ij}
\end{equation}

The average invocation time between service $i$ and $j$ can be formulated as: 
\begin{equation}
T_{ij} = D_{{S}}^{ij} + T_{S}^j
\end{equation}
A service instance can invoke external services sequentially or parallelly. We define a set services that is invoked strictly sequentially by a service group, therefore the total invocation time for a service group \( i \) is formulated as:
\begin{equation}
T_{G}^{i} = \sum_{j \in g_{i}} T_{ij}
\end{equation}

The average service processing time \( T_{S}^i \) for service \( i \) is recursively defined as the sum of its own internal execution time \( T_{E}^{i} \) and the maximum total invocation time among all service groups \( j \) that are invoked by service \( i \):

\begin{equation}
T_{S}^i = T_{E}^{i} + \max_{j \in \omega_i} T_{G}^{j}
\end{equation}

We define the first service with which the end user interacts as the gateway service \( g \) of the MSA application. The average latency between end users and the gateway service, denoted as \( D_{\text{gateway}} \) is calculated by averaging the latencies of end users to each instance of the gateway service.

\begin{equation}
D_{{gateway}} = \frac{1}{S_{{cur}}^g} \sum_{i \in \psi_g}  D_{{user}}^{i}
\end{equation}

The end-to-end latency \( D_{{msa}}\) of the MSA application can then be calculated by adding the average latency between the end user \(D_{{gateway}} \) and the average service processing time of the gateway service \( T_{{S}}^g\), which recursively includes the processing times of all subsequent services:
\begin{equation}
 D_{{msa}} = D_{{gateway}} + T_{S}^g
\end{equation}
The primary objective of our rescheduling algorithm is to optimize the end-to-end latency \( D_{{msa}} \) using the minimal rescheduling steps. This challenge, often referred to as the service placement problem, involves finding the placement of services to meet specific performance goals. However, the problem is NP-hard \cite{gu2021layer}, making optimal solutions computationally impractical. Consequently, we employ RL to derive approximate placement solutions that aim to minimize \( D_{{msa}} \).



\section{Reinforcement Learning (RL) Model}
\label{title:RL-model}
Reinforcement Learning (RL) is an adaptive decision-making paradigm where agents iteratively learn policies through interactions with the environment and reward feedback. We develop \textbf{CEEnv}, a cloud-edge continuum simulation environment, as detailed in Section~\ref{title:pipeline}. The RL agent is trained in CEEnv prior to deployment on the real-world testbed.

\begin{figure}
    \centering
    \includegraphics[width=0.70\linewidth]{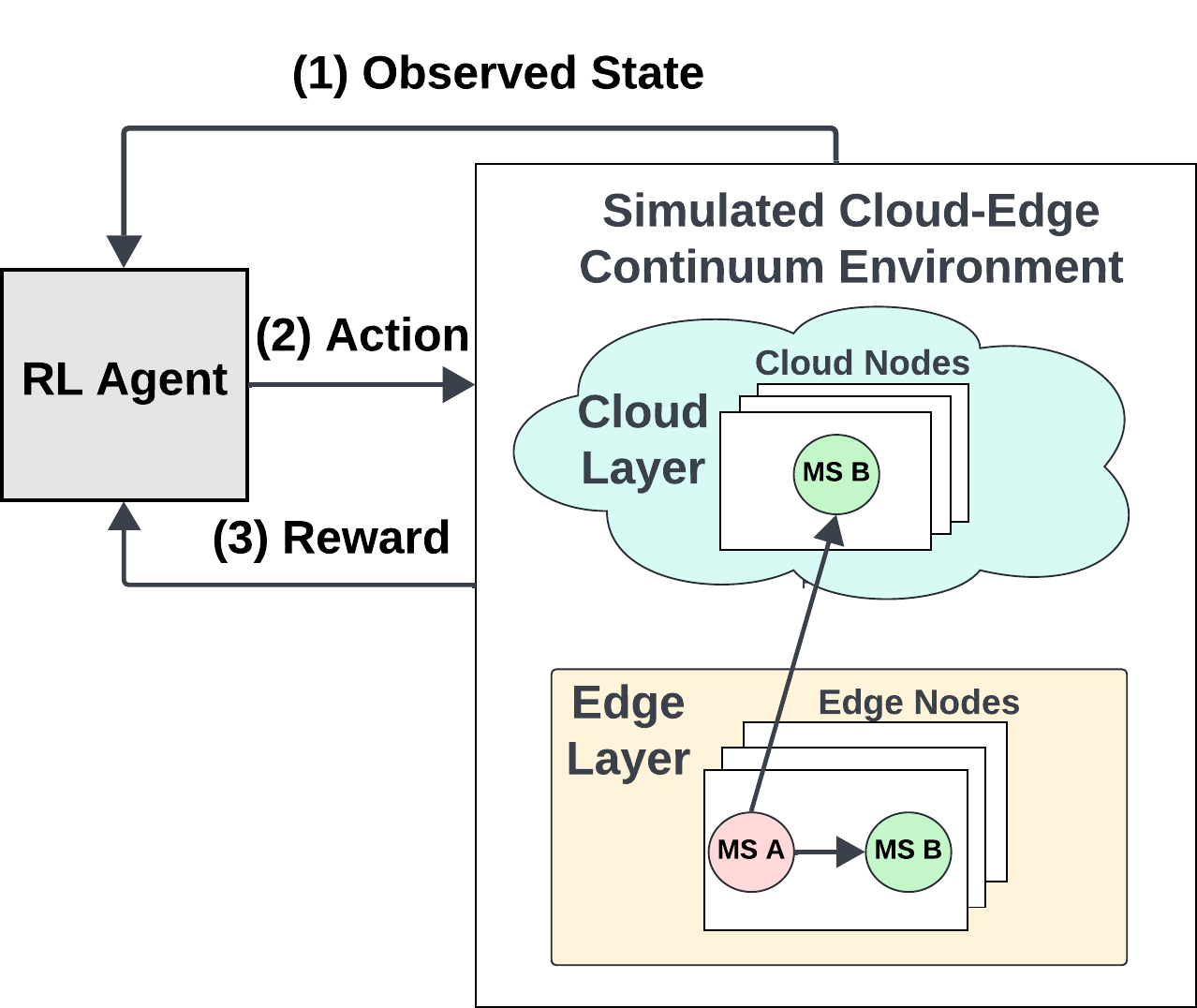}
    \caption{Interaction Between RL Agent And Environment}
    \label{fig:rl-interaction}
    \vspace{-0.5cm}
\end{figure}

The interaction between the RL agent and CEEnv is illustrated in Figure~\ref{fig:rl-interaction}. At each decision step, the agent observes the current state \(S_t\), which captures the cloud--edge continuum and the status of service instances in the MSA application. Based on its learned policy \(\pi(A_t|S_t)\), the agent generates a rescheduling action \(A_t\), selecting a service instance and assigning it to a target node. Executing this action may alter the application's end-to-end latency, which is quantified by the reward function. The agent then receives the updated state \(S_{t+1}\). The following summarizes the key components of the proposed reinforcement learning model.

\textbf{Agent:} We employ Proximal Policy Optimization (PPO) to address the rescheduling problem. Compared to value-based RL algorithms, PPO is selected for its clipping mechanism, which limits changes in the policy \( \pi(a|s; \theta^\pi) \). This constraint prevents large policy shifts that destabilize the learning process, a critical property for developing robust policy in highly dynamic cloud-edge environments.




\textbf{Episode:} At the beginning of each training sequence, the simulation environment initializes the computing resources of the cloud-edge continuum with randomly selected CPU and memory utilization values. The MSA application is then randomly deployed across these resources. From this initial state, the RL agent interacts with the environment continuously until it selects the "idle" action or reaches the maximum step limit for a single episode.

\textbf{State:} Providing informative states is crucial for enabling the RL agent to learn effective rescheduling policies in the cloud--edge continuum. 
For computing resources, the current availability of CPU and memory at each node is represented as vectors  \(\mathbf{A}_{cpu} = (A_{cpu}^1, \dots, A_{cpu}^n)\) and  \(\mathbf{A}_{mem} = (A_{mem}^1, \dots, A_{mem}^n)\),  constrained by their respective capacities  \(\mathbf{C}_{cpu} = (C_{cpu}^1, \dots, C_{cpu}^n)\) and  \(\mathbf{C}_{mem} = (C_{mem}^1, \dots, C_{mem}^n)\).  For each pod in the MSA application, CPU and memory requests are denoted by  \(\mathbf{r}_{cpu} = (r_{cpu}^1, \dots, r_{cpu}^n)\) and  \(\mathbf{r}_{mem} = (r_{mem}^1, \dots, r_{mem}^n)\).  The deployment of pods across nodes is represented by the vector  \(\mathbf{n} = (n_1, \dots, n_n)\).

\textbf{Action Space: }The RL agent's action is to select a service instance \(i\) from the current MSA application and rescheduling it to a target node \(j\), denoted as an action pair \(A(i, j)\). Furthermore, the RL agent can choose an \textbf{Idle} action, which indicates no rescheduling, thereby avoiding unnecessary movements when the current placement is already optimal. The set of possible actions can be formally defined as:
\begin{equation}
A =
\begin{cases}
A(i, j)\     i\in \varepsilon,  j\in \tau\\
{Idle}\\
\end{cases}
\end{equation}
The total number of possible rescheduling actions is calculated as \( S_{{max}} \times N + 1 \) including the idle action, where \( S_{{max}} \) denotes the maximum number of service instances in MSA applications and \( N \) represents the total number of nodes in the cloud-edge continuum.

We use the masking mechanism of Huang et al. \cite{huangCloserLookInvalid2022} to prevent the selection of invalid actions by adjusting the logits during training and inference, thus accelerating convergence.

\textbf{Reward Design:} The RL agent learns policies through a sequence of rescheduling operations aimed at minimizing end-to-end application latency. After each rescheduling step, the agent immediately receives feedback from the simulation environment in terms of latency variation. Specifically, the reward is formulated as:
\begin{equation}
\label{eq:reward}
Reward = (D_{msa}^{before} - D_{msa}^{after}) + Penalty_{cost},
\end{equation}
Where \(D_{msa}^{before}\) and \(D_{msa}^{after}\) denote the latencies before and after a rescheduling action, respectively, and \(Penalty_{cost}\) represents the overhead introduced by redeployment. 

In our training process, we found this simple reward design to be surprisingly effective. The stepwise feedback stabilizes training by mitigating the credit assignment problem\cite{suttonTemporalCreditAssignment1984}, in contrast to one-time microservice scheduling where the agent generates an entire scheduling plan and only receives rewards after full execution. It is important to note that the penalty term must be carefully scaled relative to the latency reduction; if set too high, the agent may prematurely favor idle actions, thereby limiting exploration. In our setup, where latency ranges from 100 ms to 600 ms, empirical results show that setting \(Penalty_{cost}=2\) yields effective policies while minimizing unnecessary rescheduling.

\begin{figure}[H]
    \centering
    \includegraphics[width=1\linewidth]{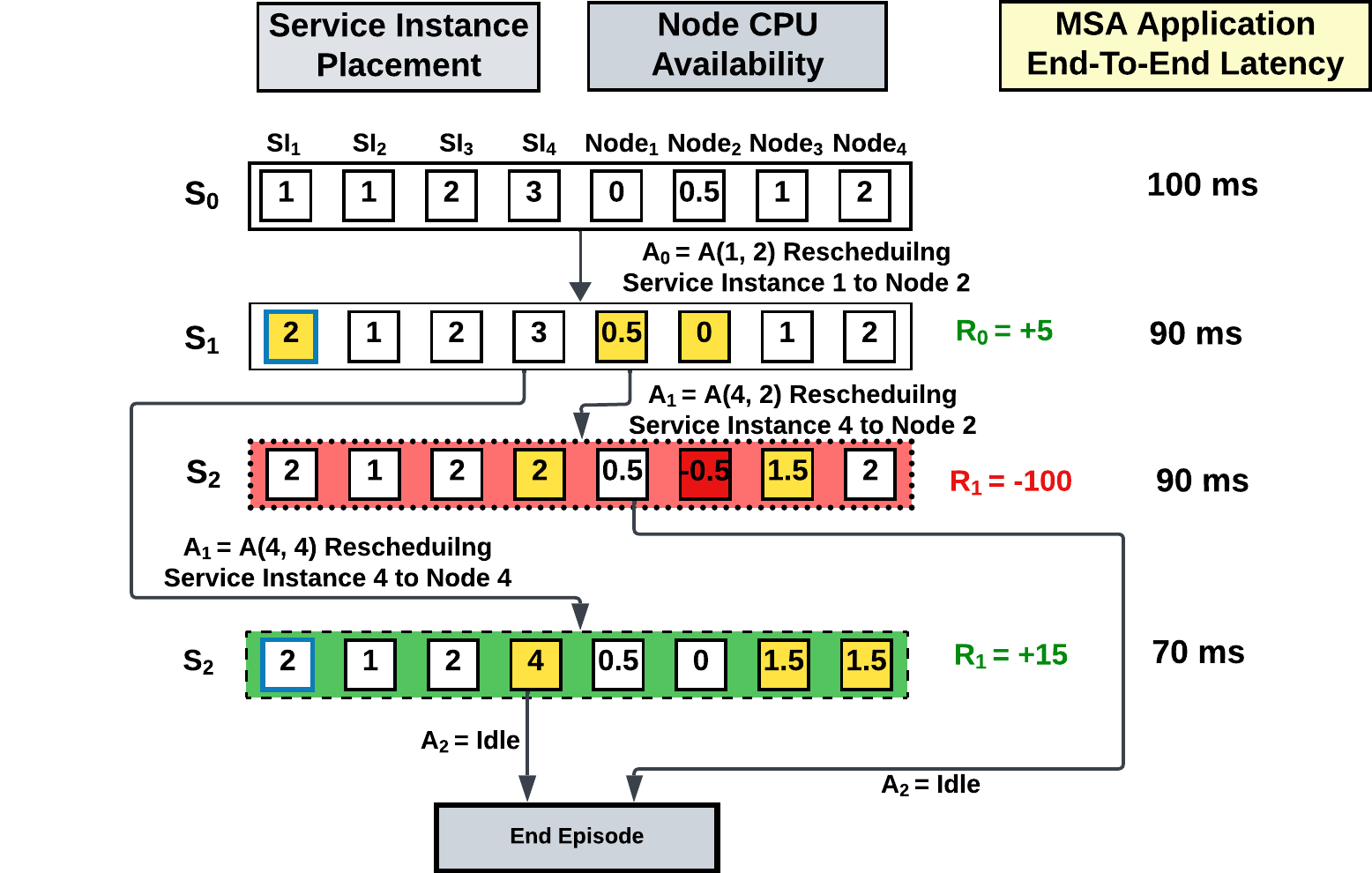}
    \caption{Example of State Transitions in Proposed RL Model}
    \label{fig:rl-workout}
\end{figure}

\textbf{Example of Episodes for the Proposed RL Model:} Figure~\ref{fig:rl-workout} shows two episodes of the RL agent interacting with the environment, focusing on pod placement and CPU availability of nodes. In both episodes, four service instances \{$SI_1$, $SI_2$, $SI_3$, $SI_4$\} are deployed across four nodes \{$Node_1$, $Node_2$, $Node_3$, $Node_4$\}, each requesting 0.5 CPU units. Initially, $SI_1$ and $SI_2$ are in $Node_2$, while $SI_3$ and $SI_4$ are on $Node_3$ and $Node_4$. The starting latency is 100 ms.

In the first episode, the agent performs an action \( A_0 = A(1,2) \), rescheduling $SI_1$ to $Node_2$. This increases $Node_1$'s CPU to 0.5 and decreases $Node_2$'s to 0, reducing latency to 90 ms. The reward is computed as \( 100 - 90 - 5 = 5 \). Next, an invalid action \( A_1 = A(4,2) \) is taken, leading to insufficient resources on $Node_2$. This ends the episode with a penalty of \( -100 \). In the second episode, the agent starts similarly, reaching state $S_1$. Then it performs \( A_1 = A(4,3) \), moving $SI_4$ to $Node_4$, reducing latency to 70 ms, and earning a reward of \( 90 - 70 - 5 = 15 \). The episode ends when the agent chooses an ``Idle" action based on system state $S_2$.

\section{Reinforcement Learning Deployment Pipeline}
\label{title:pipeline}
\label{title:profiler}
\label{title:k8s-rescheduling}

To facilitate the transfer of reinforcement learning policies from the simulation environment to real-world deployment, we design a three-stage deployment pipeline that aligns the simulation environment with real-world conditions.

\textbf{Microservice Application Profiles}: During the profiling phase, we collect performance metrics. Each microservice in the target application is iteratively deployed on nodes with varying computational resources, and its average service execution time on each node is recorded. These profiling data are subsequently streamlined into the RL simulation environment and serve as input parameters to guide environment modeling and policy training.

\textbf{Simulated Environment for Reinforcement Learning:} The reward signal is central to guiding an RL agent toward an optimal policy. To support this, we implement CEEnv, a simulation environment that closely mirrors a real-world cloud-edge continuum testbed. In our reward design (Section~\ref{title:RL-model}), variations in the end-to-end latency of the MSA are the primary factor; thus, CEEnv focuses on accurately modeling this latency. For a given application, CEEnv takes as input a microservice invocation graph (JSON) and profiled execution times across heterogeneous devices. During training, it traverses the invocation graph via DFS, recursively aggregating execution times and inter-service latency (Section~\ref{title:problem-formulation}).

When microservices include multiple external calls, CEEnv distinguishes between parallel and sequential patterns (Section~\ref{title:system-model}) to compute their impact. By combining application models, profiling data, and cloud--edge dynamics, CEEnv yields latency simulations consistent with real-world setups, enabling policies learned in simulation to transfer effectively to physical testbeds.

\textbf{Integrating with Kubernetes}:
After training on CEEnv, we integrate the trained rescheduling policy into Kubernetes. Since Kubernetes lacks native support for pod rescheduling, we developed a custom plugin for its scheduler that continuously monitors the cluster status and running MSA applications, performing rescheduling when necessary. During rescheduling, new pods are launched on the target nodes, and the old pods are terminated only after the new ones are running, thereby minimizing the impact on service availability.
\vspace{-1.5em}
\section{Experimental Settings}
\subsection{\textbf{Testbed Setup}}
We utilize Kubernetes to deploy an cloud-edge continuum testbed on virtual machines (VMs) provisioned by the University OpenStack~\footnote{https://www.openstack.org/} Cloud Computing Platform. The VM specifications are summarized in Table~\ref{tab:eval-vm-spec}.

We configure the testbed to capture both network characteristics and resource heterogeneity in the cloud–edge continuum. Cloud and edge resources are deployed with an average latency of 50 ms, while the raw network latency between edge and client nodes remains below 1 ms. In addition, all nodes within the same domain (cloud or edge) are interconnected with an average latency below 1 ms.

To approximate the heterogeneity of real-world pervasive computing environments, we incorporate different types of computing resources with varying performance levels that directly influence task execution times. For example, services running on Cloud-A nodes execute on average 1.5× faster than on Edge-A nodes and 2× faster than on Edge-B nodes.

\vspace{-0.5em}

\begin{table}[H]
\caption{TESTBED VIRTUAL MACHINES SPECIFICATIONS}
\label{tab:eval-vm-spec}
\resizebox{\columnwidth}{!}{%
\begin{tabular}{llllll}
\hline
VM Type & CPU Cores & Memory & Storage & Operating System & Count \\ \hline
Cloud-A & 8 & 16GB & 30GB & Ubuntu 22.04 & 2 \\
Edge-A & 4 & 8GB & 30GB & Ubuntu 22.04 & 2 \\
Edge-B & 2 & 4GB & 30GB & Ubuntu 22.04 & 2 \\
Client-A & 4 & 8GB & 30GB & Ubuntu 22.04 & 1 \\ \hline
\vspace{-1.5em}
\end{tabular}%
}
\end{table}

\subsection{\textbf{Workload}}
\label{title:MSA-application}
\begin{figure}[H]

    \centering
    \includegraphics[width=1\linewidth]{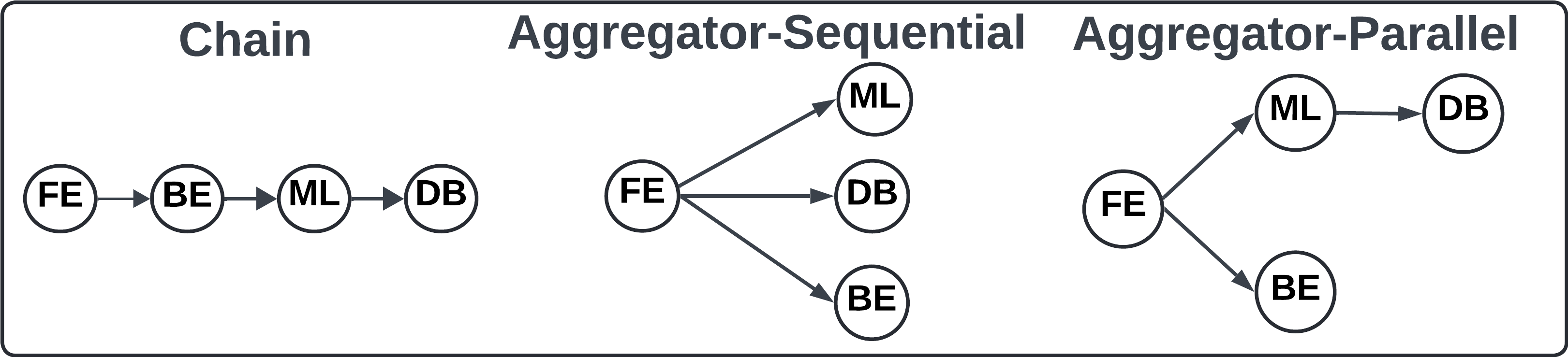}
    \caption{Constructed benchmark MSA applications with varying service invocation patterns and orders.}
    \label{fig:imple-msa}
\end{figure}

The test our approach on MSA applications with different invocation patterns, we employ \texttt{\(\mu\text{Bench}\)} \cite{dettiMBenchOpenSourceFactory2023b} to build three microservice benchmark applications, as shown in Figure~\ref{fig:imple-msa}. The \textbf{Chain} application features a linear pattern where each service calls another service. The \textbf{Aggregator-Sequential} application involves a front-end service that sequentially invokes multiple services, while the \textbf{Aggregator-Parallel} application allows the front-end to call two services concurrently. All applications use the same microservices, but differ in their interaction patterns, highlighting various invocation structures.
Each MSA application comprises four services, each service supporting up to five replicas, resulting in a maximum of 20 pods per application.

A microservice configured through \texttt{\(\mu\text{Bench}\)} can stress CPU resources by calculating digits of \(\pi\) up to a specified complexity, stress memory by loading a specific amount of data, and stress disk by writing a defined number of bytes. Detailed parameters for the usage of resources for each microservice are provided in Table~\ref{table:ms-config}.

\begin{table}[H]
\caption{MICROSERVICE APPLICATION CONFIGURATION}
\label{table:ms-config}
\resizebox{\columnwidth}{!}{%
\begin{tabular}{|l|l|l|l|l|l|}
\hline
 & Client & Front-End & ml & Back-End & DB \\ \hline
Pi calculation complexity & 1 & 200 & 400 & 200 & 100 \\ \hline
Iteration of pi calculation & 1 & 2 & 8 & 4 & 1 \\ \hline
Memory writing bytes (KB) & 0 & 1000 & 2000 & 1000 & 100 \\ \hline
Disk writing & 0 & 1 & 1 & 1 & 100 \\ \hline
\end{tabular}%
}
\end{table}

\vspace{-0.5em}

\subsection{\textbf{Metrics}}
\label{title:eval-metrics}
In this work, we implement a Go-based workload generator for MSA applications, using goroutines to dispatch requests sequentially upon receiving prior responses. The intensity of the work load is controlled by adjusting the number of goroutines. The proposed rescheduling algorithms are evaluated using \textbf{End-to-end latency} which measures the total time from the time user requests are initiated to when responses are received.

\subsection{\textbf{Baseline Algorithms}}
\label{title:eval-baseline}
We evaluate two types of algorithms: \textbf{scheduling} algorithms for initial placements of MSA applications and \textbf{rescheduling} algorithms for optimizing existing placements. We benchmark proposed REACH rescheduling algorithm against three scheduling and one rescheduling baseline.

\textbf{Default}: The default algorithm of the Kubernetes Scheduler\footnote{https://kubernetes.io/docs/concepts/scheduling-eviction/kube-scheduler/}, which assigns pods to nodes that have sufficient computing resources to meet the resource requirements of pods.

\textbf{Cloud-First}: Prioritizes offloading tasks to cloud nodes, alleviating resource constraints on edge devices. This approach takes advantage of the greater computational capacity of the cloud to handle intensive workloads.

\textbf{Latency-Greedy\cite{mota-cruzOptimizingMicroservicesPlacement2024}}: Using latency from service to end user as heuristic, it focuses on minimizing latency by keeping computation near data sources or end-users.

\textbf{RSDQL\cite{lvMicroserviceDeploymentEdge2022a}}: Employs the Deep Q-Learning reinforcement learning method with a replay buffer mechanism to tackle microservice application scheduling in the cloud-edge environment. As an initial placement policy, it demonstrates superior performance over heuristic algorithms. We adapt RSDQL algorithm to our rescheduling model to serves as a baseline for evaluating our proposed method.

\section{Performance Evaluation}
In this section, we first analyze the convergence behavior and scalability of our RL algorithm. We then present a performance evaluation of REACH on the real-world testbed, focusing on its effectiveness in reducing application latency and handling dynamic environmental conditions.



\begin{table}[H]
\caption{Hyper-Parameters For RL Agents}
\label{tab:hyperparameters}
\resizebox{\columnwidth}{!}{%
\begin{tabular}{llll}
\hline
Parameter & Value & Parameter & Value \\ \hline
$Penalty_{step}$ & 2& Batch size & 64 \\
Learning Rate ($\alpha$) & $3 \times 10^{-4}$ & Exploration Fraction (RSDQL) & 0.1 \\
Discount Factor ($\gamma$) & 0.99 & \begin{tabular}[c]{@{}l@{}}No. of Fully Connected Layers \\ for Q-Network\end{tabular} & 3 \\
Clip Range ($\epsilon$) & 0.2 & \begin{tabular}[c]{@{}l@{}}No. of Fully Connected Layers \\ for Policy Network\end{tabular} & 2 \\
Entropy Coefficient ($\beta$) & 0 & \begin{tabular}[c]{@{}l@{}}No. of Fully Connected Layers \\ for Value Network\end{tabular} & 2 \\
Soft Update Coefficient ($\tau$) & 1 &  &  \\ \hline
\end{tabular}%
}
\end{table}
\begin{figure}[H]
    \centering
    \includegraphics[width=1\columnwidth]{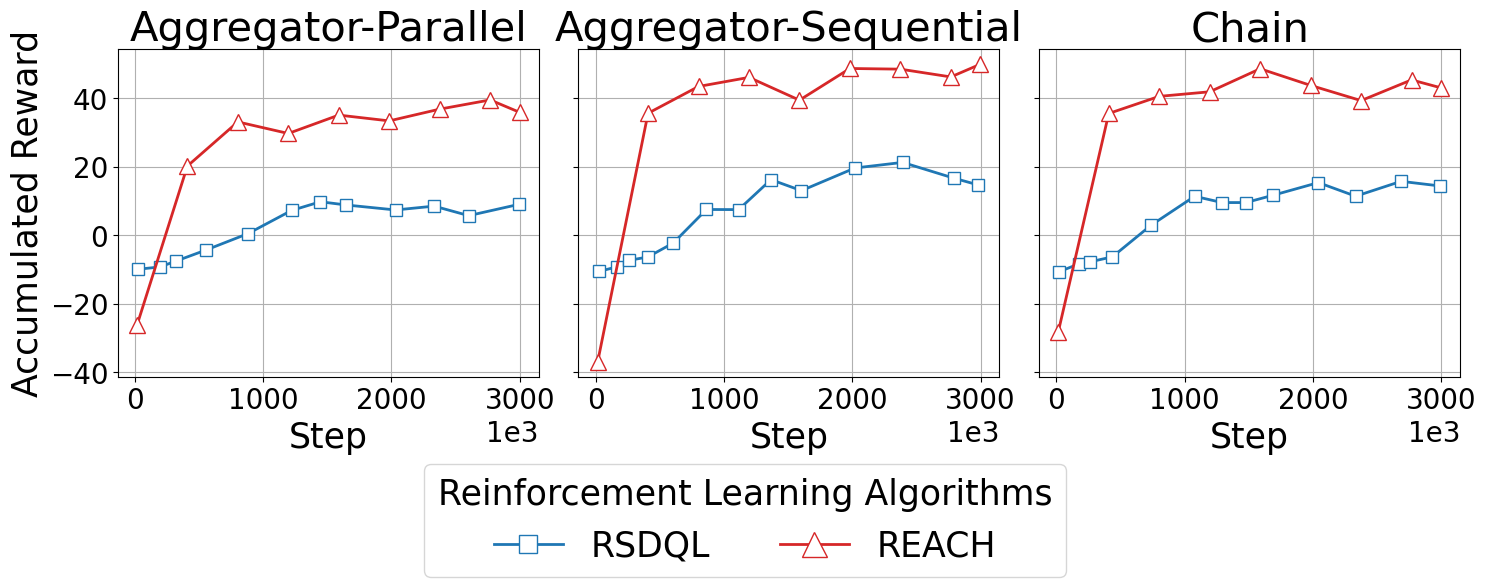}
    \caption{Convergence of accumulated rewards for REACH and RSDQL agents on Aggregator-Parallel, Aggregator-Sequential, and Chain application workloads}
    \label{fig:reward-convergence}
    
\end{figure}

\begin{figure*}[t]
    \centering
    \includegraphics[width=1\linewidth]{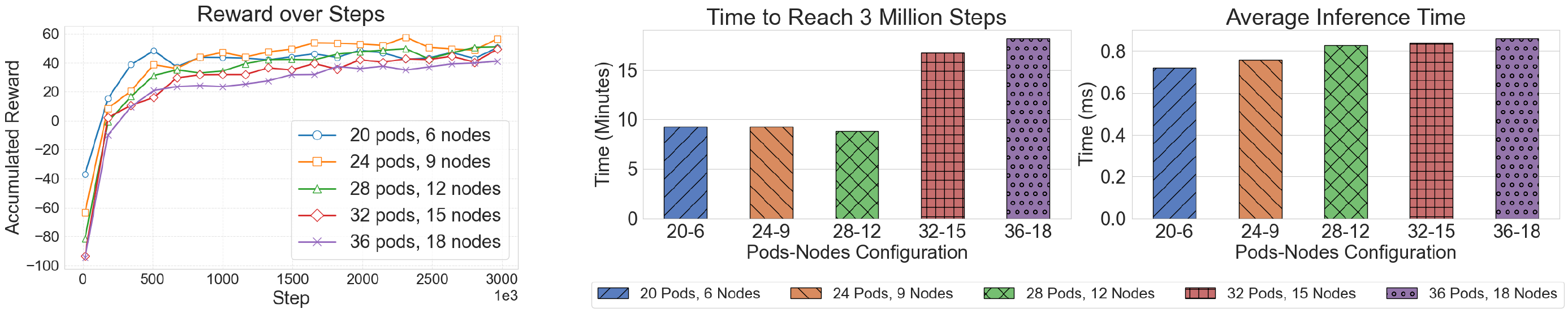}
    \caption{Training performance of REACH at larger scales, including rewards, training time, and inference time for different node and pod configurations.}
    \label{fig:scalability}
\end{figure*}

\vspace{-1em}
\subsection{\textbf{Convergence of the RL Agents}}
\label{title:eval-convergence}
In this study, we trained REACH and RSDQL agents on a virtual machine equipped with 64 cores AMD 9474F CPU, 128G of memory and L40s GPU. The hyperparameters for both REACH and RSDQL are listed in Table~\ref{tab:hyperparameters}, it includes both training and reward parameters from our reward function. We limit the training steps to 3 million. 

Figure~\ref{fig:reward-convergence} illustrates the accumulated rewards per episode for REACH and RSDQL across three MSA applications, averaged over 100 episodes. Both models were trained in a simulated environment with six computing nodes and up to 20 pods. REACH achieves higher cumulative rewards, indicating a more effective rescheduling policy, owing to the stability and sample efficiency of the PPO algorithm it employs.

The RL models above require less than 10 minutes of training, in contrast to direct training on a real-world testbed, which may take several days. Leveraging the simulation environment, our approach enables significantly more efficient training while ensuring that the learned policies can be effectively transferred and applied to the real-world cloud–edge continuum.

\subsection{\textbf{Scalability Analysis}}
\label{title:eval-scalability}

We evaluated REACH’s scalability by enlarging the simulated cloud–edge continuum with additional computing nodes and application pods. We choose Aggregator–Sequential application with the longest training time over others, experiment results are summarized in Figure~\ref{fig:scalability}. As pods increased from 20 to 36 and nodes from 6 to 18, rewards across scales converged rapidly above 20, with larger configurations showing  slower learning. Training time grew with system size yet remained under 20 minutes. Meanwhile, policy inference latency per decision step stayed nearly constant (within 1~ms) across scales, indicating negligible online scheduling overhead and supporting real-time deployment in larger clusters.

\begin{figure}[t]
    \centering
    \includegraphics[width=0.85\linewidth]{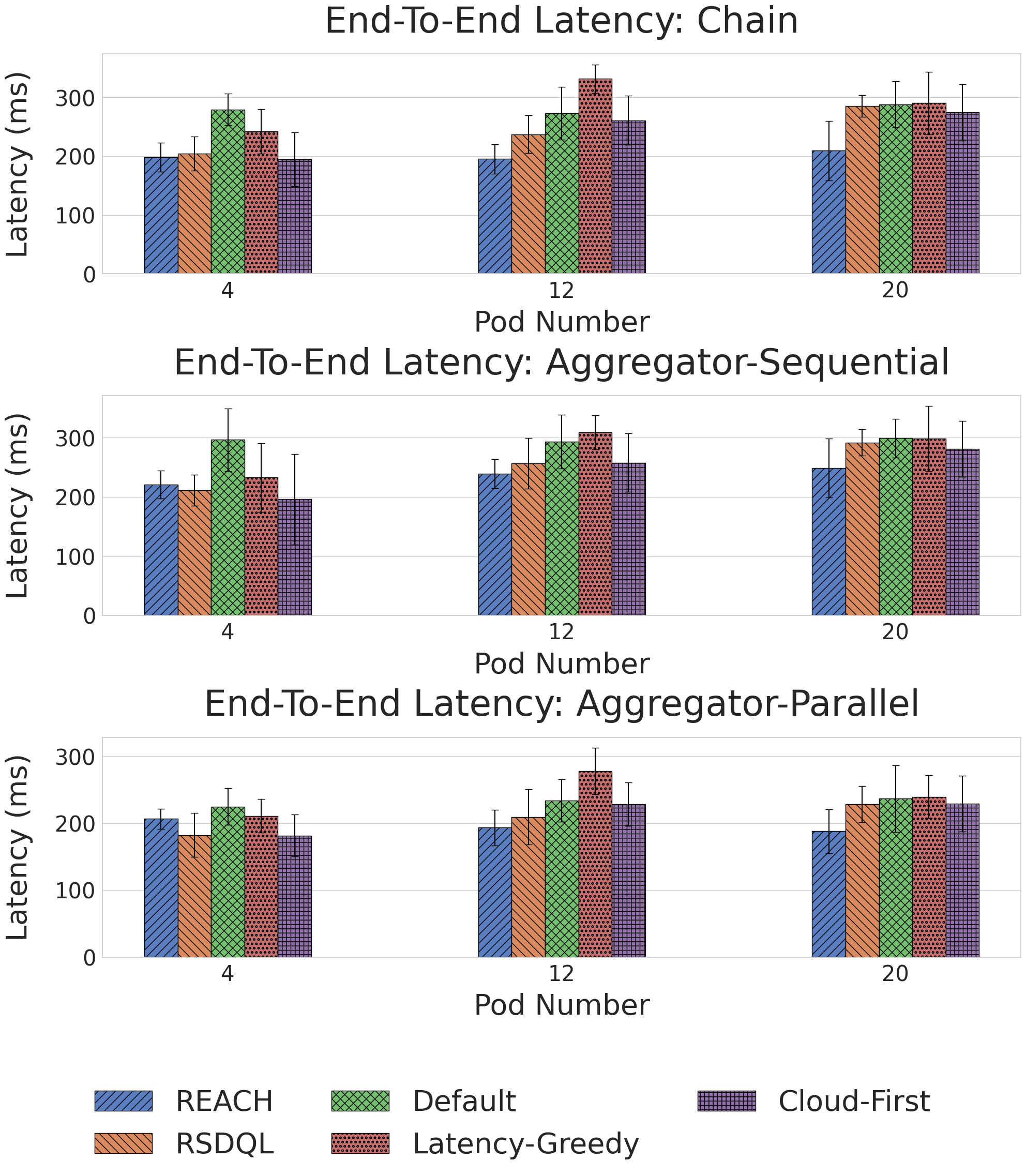}
    \caption{Comparison of end-to-end latency across three MSA applications with different pod numbers.}
    \label{fig:eval-quan-e2e}
    \vspace{-0.5em}
\end{figure}

\vspace{-0.5em}
\subsection{\textbf{Evaluation of End-to-End Latency}}
\label{title:eval-e2e}
We evaluated end-to-end latency in three MSA applications scheduled by different algorithms, as shown in Figure~\ref{fig:eval-quan-e2e}. Each application was tested with pod configurations of 4, 12, and 20, achieved by setting service replicas to 1, 3, and 5, respectively. 

REACH outperformed the baselines, reducing latency by an average of $21.56\%$ and $9.83\%$ compared to default and RSDQL respectively. REACH demonstrated superior performance in the 12 and 20-pod configurations, but was outperformed by RSDQL and Cloud-First in the 4-pod setup of Aggregator-Sequential and Aggregator-Parallel.

As discussed in Section~\ref{title:system-model}, our network and MSA models introduce certain abstractions from the real-world testbed, which may lead to minor discrepancies between simulated and real environments. Nevertheless, REACH outperforms other algorithms in 7 out of 9 pod configurations with more complex setups and shows only a small gap to RSDQL in the 4-pod configuration, highlighting its balance between model efficiency and performance in a real-world testbed.





\vspace{-0.5em}
\subsection{\textbf{Evaluation of Adaptability Under Cluster Dynamics}}
We designed a node failure scenario to assess the adaptability of our proposed rescheduling algorithm under dynamic cluster changes. The experiment begins by deploying the MSA application, where each service runs with three replicas, and the traffic generator operates with a single thread, generating traffic in a best-effort manner.

\begin{figure*}
    \centering
    \includegraphics[width=1\linewidth]{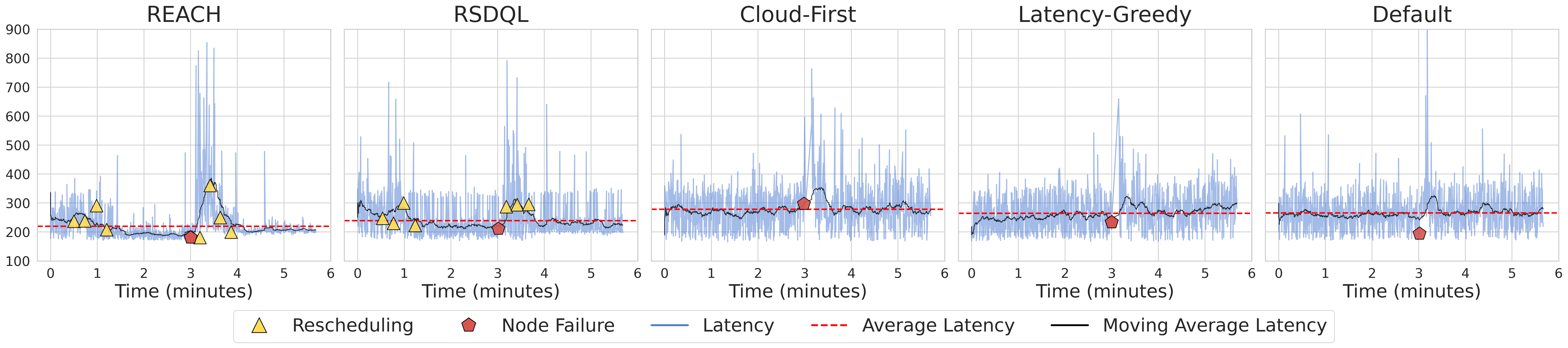}
    \caption{Latency trends for the Chain Application across algorithms in the node failure case. Rescheduling and node failure events are represented as markers in the graph. REACH outperforms all other algorithms overall, effectively reducing fluctuations and eliminating spikes.}
    \label{fig:ts-nodefailure}
    \vspace{-0.2cm}
\end{figure*}

At a predefined time, we simulate node failures by disabling one Edge-A and one Cloud-A node in the testbed. The Kubernetes self-healing mechanism restores the affected pods to the remaining nodes. For the heuristic algorithm, scheduling actions are performed during the self-healing process, while REACH and RSDQL perform further pod rescheduling after the self-healing process.

Figure~\ref{fig:ts-nodefailure} shows end-to-end latency trends for the Chain application. Each subgraph corresponds to one algorithm, with node failures marked by orange pentagons and rescheduling actions by yellow triangles. The moving average (window size 30) highlights overall trends, and the red dashed line indicates the average latency as a baseline.

Results demonstrate that REACH reduces latency fluctuations more effectively than other algorithms, maintaining a smoother curve with fewer spikes after rescheduling. REACH keeps 86.9\% of requests below 250 ms, compared to 76.9\% for RSDQL and much lower proportions for Cloud-First (35.5\%), Latency-Greedy (46.2\%), and Default (46.7\%). Its average latency of 219.92 ms represents improvements of 7.9\% over RSDQL, 24.5\% over Cloud-First, 18.7\% over Latency-Greedy, and 19.1\% over Default.




\begin{figure*}
    \centering
        \includegraphics[width=1\linewidth]{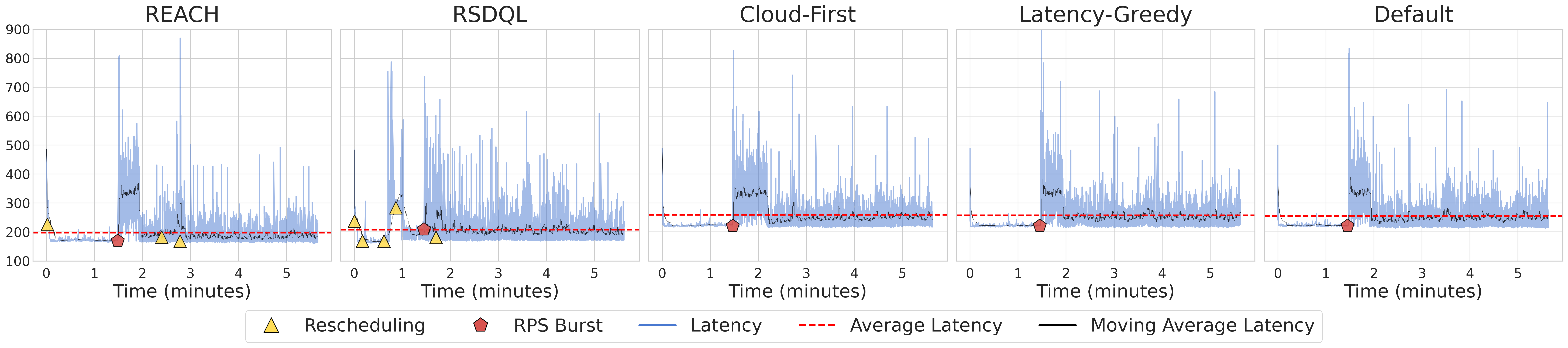}
    \caption{Latency trends for the Chain Application across algorithms when rescheduling works with autoscaling. A traffic surge is observed at 90 seconds. REACH outperforms all other algorithms by reducing overall latency and minimizing fluctuations.}
    \label{fig:ts-autoscaling}
    \vspace{-0.2cm}
\end{figure*}
\begin{figure*}
    \centering
        \includegraphics[width=1\linewidth]{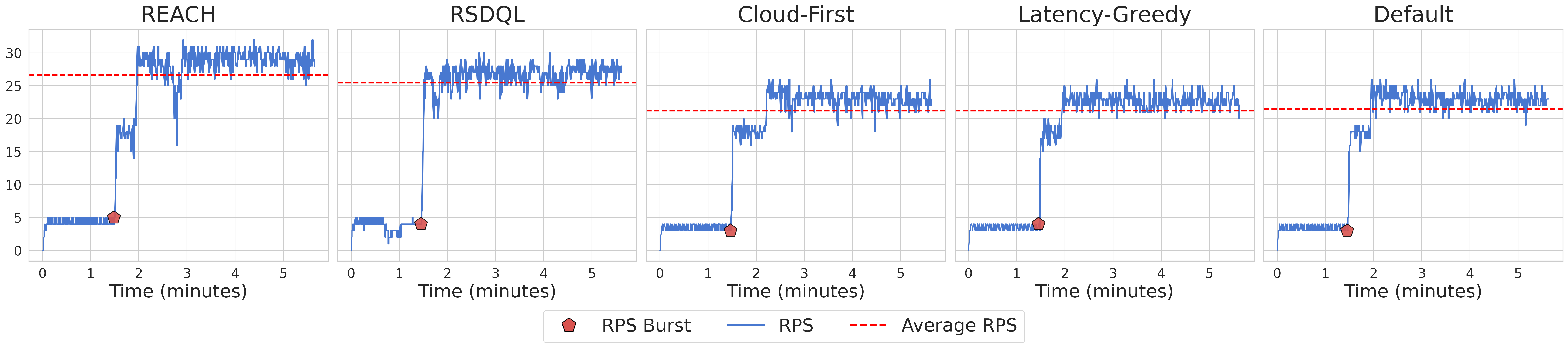}
    \caption{Comparison of the RPS handled by the Chain application over time across different algorithms. The application optimized by REACH sustains a higher request rate than other baseline algorithms.}
    \label{fig:ts-rps}
\end{figure*}

We conducted the same experiment on other two applications. In these scenarios, REACH achieves average latencies of 190.25 ms and 220.1 ms, outperforming the best baselines by over 10\% and 8\%, respectively. 

\vspace{-0.1em}
\subsection{\textbf{Evaluation of Adaptability to User Request Variability}}
\label{title:adaptabiliy}
We designed an experiment to evaluate our rescheduling algorithm with dynamically changing user requests. The Kubernetes autoscaling mechanism was enabled with a CPU usage threshold of 50\%. 

The workload generator began with one thread and increased to six to emulate a surge in requests. During the experiment, heuristic algorithms managed initial pod placement and autoscaling, while REACH and RSDQL performed rescheduling. Figure~\ref{fig:ts-autoscaling} shows latency trends, with surges marked by orange pentagons. REACH achieved the lowest average latency of 197.55 ms, outperforming RSDQL (207.29 ms), Cloud-First (258.77 ms), Latency-Greedy (258.02 ms), and Default (255.12 ms). It also exhibited greater stability, keeping 90\% of requests under 250 ms compared to 89\% for RSDQL and only 55–59\% for the other baselines.  

Figure~\ref{fig:ts-rps} presents throughput results. Operating in best-effort mode, the workload generator issues a new request upon completion of the previous one; thus, lower response times yield higher RPS. After the surge, REACH sustained the highest throughput, averaging 26.62 RPS versus 25.45 (RSDQL) and about 21 for the other baselines.  

We repeated the experiment on the Aggregator-Parallel and Aggregator-Sequential applications, where REACH further reduced average latency by 14.48\% and 3\% over the best baseline, respectively, while also mitigating traffic fluctuations and maintaining higher throughput.

\subsection{\textbf{Rescheduling Overhead}}

The number of rescheduling steps is a critical overhead when adjusting MSA application placements in response to system changes. During the end-to-end latency experiments (Section~\ref{title:eval-e2e}), the REACH algorithm required an average of 1.58, 6.2, and 8.95 rescheduling actions for the 4-, 12-, and 20-pod configurations, respectively. Unlike conventional scheduling approaches that often reschedule the entire application, only 39.5\%, 52\%, and 42\% of active pods were migrated on average across these configurations, effectively reducing unnecessary overhead.

Another source of overhead is continuous ``pod churn'' from frequent rescheduling, which occurs when algorithms repeatedly attempt to restore latency below an unattainable threshold under changed conditions. Our RL approach mitigates this by modeling rescheduling overhead with a soft threshold, triggering actions only when expected benefits exceed penalties. This promotes stable policies that avoid self-looping and overreaction without relying on rigid latency targets. In both node failure and auto-scaling scenarios (Section~\ref{title:adaptabiliy}), REACH executed only 4 and 2 rescheduling actions, respectively, effectively avoiding unnecessary adjustments even when original performance could not be fully restored.

Lastly, overhead may also arise from transient latency spikes during rescheduling. As noted in Section~\ref{title:k8s-rescheduling}, our implementation avoids hard service interruptions when new pods are launched. However, brief spikes typically under 1 second can occur as Kubernetes redirects traffic from old pods. Given their short duration and the substantial overall latency reduction achieved, we consider this overhead acceptable.



\section{Conclusions and Future Work}

In this paper, we present a novel rescheduling algorithm to optimize microservice placement in hybrid cloud–edge environments. Unlike traditional methods that reschedule entire applications, our approach selectively updates only necessary pods in real time, reducing overhead while maintaining continuous performance. Reinforcement learning is employed to adapt placements based on real-time invocation patterns and resource availability across the continuum.  

Experiments on a real-world Kubernetes testbed show that our algorithm significantly reduces end-to-end latency and mitigates fluctuations and spikes caused by dynamic cluster changes and varying workloads, outperforming baselines. 

Furthermore, our work addresses the challenge of integrating RL-based approach in real-world cloud–edge environments, though some limitations remain. Our approach can be extended to other objectives such as energy efficiency and bandwidth optimization. Due to resource constraints, our testbed is limited in scale. Nevertheless, since our system is built on Kubernetes, it can be easily scaled to larger deployments as resources become available.

For future work, we plan to extend the RL framework toward multi-objective optimization, incorporating energy and bandwidth metrics. We also aim to combine simulation pre-training with online learning in real testbeds to further enhance adaptability and performance.

\bibliography{bibliography}

\begin{thebibliography}{10}
\providecommand{\url}[1]{#1}
\csname url@samestyle\endcsname
\providecommand{\newblock}{\relax}
\providecommand{\bibinfo}[2]{#2}
\providecommand{\BIBentrySTDinterwordspacing}{\spaceskip=0pt\relax}
\providecommand{\BIBentryALTinterwordstretchfactor}{4}
\providecommand{\BIBentryALTinterwordspacing}{\spaceskip=\fontdimen2\font plus
\BIBentryALTinterwordstretchfactor\fontdimen3\font minus \fontdimen4\font\relax}
\providecommand{\BIBforeignlanguage}[2]{{%
\expandafter\ifx\csname l@#1\endcsname\relax
\typeout{** WARNING: IEEEtran.bst: No hyphenation pattern has been}%
\typeout{** loaded for the language `#1'. Using the pattern for}%
\typeout{** the default language instead.}%
\else
\language=\csname l@#1\endcsname
\fi
#2}}
\providecommand{\BIBdecl}{\relax}
\BIBdecl

\bibitem{gong2010characteristics}
C.~Gong, J.~Liu, Q.~Zhang, H.~Chen, and Z.~Gong, ``The characteristics of cloud computing,'' in \emph{2010 39th International Conference on Parallel Processing Workshops}.\hskip 1em plus 0.5em minus 0.4em\relax IEEE, 2010, pp. 275--279.

\bibitem{shi2016edge}
W.~Shi, J.~Cao, Q.~Zhang, Y.~Li, and L.~Xu, ``Edge computing: Vision and challenges,'' \emph{IEEE internet of things journal}, vol.~3, no.~5, 2016.

\bibitem{moreschini2022cloud}
S.~Moreschini, F.~Pecorelli, X.~Li, S.~Naz, D.~H{\"a}stbacka, and D.~Taibi, ``Cloud continuum: The definition,'' \emph{IEEE Access}, 2022.

\bibitem{botvinickReinforcementLearningFast2019}
M.~Botvinick, S.~Ritter, J.~X. Wang, Z.~Kurth-Nelson, C.~Blundell, and D.~Hassabis, ``Reinforcement {{Learning}}, {{Fast}} and {{Slow}},'' vol.~23, no.~5, pp. 408--422.

\bibitem{filipMicroservicesSchedulingModel2018a}
I.-D. Filip, F.~Pop, C.~Serbanescu, and C.~Choi, ``Microservices {{Scheduling Model Over Heterogeneous Cloud-Edge Environments As Support}} for {{IoT Applications}},'' \emph{IEEE Internet of Things Journal}, vol.~5, no.~4, pp. 2672--2681, Aug. 2018.

\bibitem{LightweightDecentralizedService}
C.~Guerrero, I.~Lera, and C.~Juiz, ``A lightweight decentralized service placement policy for performance optimization in fog computing,'' \emph{Journal of Ambient Intelligence and Humanized Computing}, 2019.

\bibitem{centofantiLatencyAwareKubernetesScheduling2023}
C.~Centofanti, W.~Tiberti, A.~Marotta, F.~Graziosi, and D.~Cassioli, ``Latency-{{Aware Kubernetes Scheduling}} for {{Microservices Orchestration}} at the {{Edge}},'' in \emph{2023 {{IEEE}} 9th {{International Conference}} on {{Network Softwarization}} ({{NetSoft}})}, Jun. 2023, pp. 426--431.

\bibitem{OptimalDeploymentFog}
``Optimal {{Deployment}} of {{Fog Nodes}}, {{Microservices}} and {{SDN Controllers}} in {{Time-Sensitive IoT Scenarios}} {\textbar} {{IEEE Conference Publication}} {\textbar} {{IEEE Xplore}},'' https://ieeexplore.ieee.org/abstract/document/9685995.

\bibitem{xieNovelDirectionalNonlocalconvergent2019}
Y.~Xie, Y.~Zhu, Y.~Wang, Y.~Cheng, R.~Xu, A.~S. Sani, D.~Yuan, and Y.~Yang, ``A novel directional and non-local-convergent particle swarm optimization based workflow scheduling in cloud--edge environment,'' \emph{Future Generation Computer Systems}, vol.~97, pp. 361--378, Aug. 2019.

\bibitem{alelyaniOptimizingCloudPerformance2024}
A.~Alelyani, A.~Datta, and G.~M. Hassan, ``Optimizing {{Cloud Performance}}: {{A Microservice Scheduling Strategy}} for {{Enhanced Fault-Tolerance}}, {{Reduced Network Traffic}}, and {{Lower Latency}},'' \emph{IEEE Access}, 2024.

\bibitem{mampageDeepReinforcementLearning2023a}
A.~Mampage, S.~Karunasekera, and R.~Buyya, ``A {{Deep Reinforcement Learning}} based {{Algorithm}} for {{Time}} and {{Cost Optimized Scaling}} of {{Serverless Applications}},'' Aug. 2023.

\bibitem{maDeepMultiagentReinforcement2025a}
N.~Ma, A.~Tang, Z.~Xiong, and F.~Jiang, ``A deep multi-agent reinforcement learning approach for the micro-service migration problem with affinity in the cloud,'' vol. 273, p. 126856, 2025.

\bibitem{maiaDeepReinforcementLearning2023}
A.~M. Maia and Y.~{Ghamri-Doudane}, ``A {{Deep Reinforcement Learning Approach}} for the {{Placement}} of {{Scalable Microservices}} in the {{Edge-to-Cloud Continuum}},'' in \emph{{{GLOBECOM}} 2023 - 2023 {{IEEE Global Communications Conference}}}.\hskip 1em plus 0.5em minus 0.4em\relax Kuala Lumpur, Malaysia: IEEE, Dec.

\bibitem{chenIoTMicroserviceDeployment2021}
L.~Chen, Y.~Xu, Z.~Lu, J.~Wu, K.~Gai, P.~C.~K. Hung, and M.~Qiu, ``{{IoT Microservice Deployment}} in {{Edge-Cloud Hybrid Environment Using Reinforcement Learning}},'' \emph{IEEE Internet of Things Journal}, vol.~8, no.~16, pp. 12\,610--12\,622, Aug. 2021.

\bibitem{lvGraphReinforcementLearningBasedDependencyAwareMicroservice2024}
W.~Lv, P.~Yang, T.~Zheng, C.~Lin, Z.~Wang, M.~Deng, and Q.~Wang, ``Graph-{{Reinforcement-Learning-Based Dependency-Aware Microservice Deployment}} in {{Edge Computing}},'' \emph{IEEE Internet of Things Journal}, vol.~11, no.~1, pp. 1604--1615, Jan. 2024.

\bibitem{afachaoEfficientMicroserviceDeployment2024}
K.~Afachao, A.~M. Abu-Mahfouz, and G.~P. Hanke, ``Efficient {{Microservice Deployment}} in the {{Edge-Cloud Networks}} with {{Policy-Gradient Reinforcement Learning}},'' pp. 1--1.

\bibitem{fuAdaptiveResourceEfficient2022}
K.~Fu, W.~Zhang, Q.~Chen, D.~Zeng, and M.~Guo, ``Adaptive {{Resource Efficient Microservice Deployment}} in {{Cloud-Edge Continuum}},'' \emph{IEEE Transactions on Parallel and Distributed Systems}, Aug. 2022.

\bibitem{pallewattaQoSawarePlacementMicroservicesbased2022}
S.~Pallewatta, V.~Kostakos, and R.~Buyya, ``{{QoS-aware}} placement of microservices-based {{IoT}} applications in {{Fog}} computing environments,'' \emph{Future Generation Computer Systems}, vol. 131, pp. 121--136, Jun. 2022.

\bibitem{faticantiDeploymentApplicationMicroservices2020}
F.~Faticanti, M.~Savi, F.~D. Pellegrini, P.~Kochovski, V.~Stankovski, and D.~Siracusa, ``Deployment of {{Application Microservices}} in {{Multi-Domain Federated Fog Environments}},'' in \emph{2020 {{International Conference}} on {{Omni-layer Intelligent Systems}} ({{COINS}})}, Aug. 2020, pp. 1--6.

\bibitem{armaniCostEffectiveWorkloadAllocation2021}
V.~Armani, F.~Faticanti, S.~Cretti, S.~Kum, and D.~Siracusa, ``A {{Cost-Effective Workload Allocation Strategy}} for {{Cloud-Native Edge Services}},'' Oct. 2021.

\bibitem{gu2021layer}
L.~Gu, D.~Zeng, J.~Hu, B.~Li, and H.~Jin, ``Layer aware microservice placement and request scheduling at the edge,'' in \emph{IEEE INFOCOM 2021-IEEE Conference on Computer Communications}.\hskip 1em plus 0.5em minus 0.4em\relax IEEE, 2021.

\bibitem{huangCloserLookInvalid2022}
S.~Huang and S.~Onta{\~n}{\'o}n, ``A {{Closer Look}} at {{Invalid Action Masking}} in {{Policy Gradient Algorithms}},'' \emph{The International FLAIRS Conference Proceedings}, vol.~35, May 2022.

\bibitem{suttonTemporalCreditAssignment1984}
R.~S. SUTTON, ``Temporal {{Credit Assignment}} in {{Reinforcement Learning}}.''

\bibitem{dettiMBenchOpenSourceFactory2023b}
A.~Detti, L.~Funari, and L.~Petrucci, ``{{$\mu$Bench}}: {{An Open-Source Factory}} of {{Benchmark Microservice Applications}},'' \emph{IEEE Transactions on Parallel and Distributed Systems}, vol.~34, no.~3, pp. 968--980, Mar. 2023.

\bibitem{mota-cruzOptimizingMicroservicesPlacement2024}
M.~Mota-Cruz, J.~H. Santos, J.~F. Macedo, K.~Velasquez, and D.~P. Abreu. Optimizing {{Microservices Placement}} in the {{Cloud-to-Edge Continuum}}: {{A Comparative Analysis}} of {{App}} and {{Service Based Approaches}}.

\bibitem{lvMicroserviceDeploymentEdge2022a}
W.~Lv, Q.~Wang, P.~Yang, Y.~Ding, B.~Yi, Z.~Wang, and C.~Lin, ``Microservice {{Deployment}} in {{Edge Computing Based}} on {{Deep Q Learning}},'' \emph{IEEE Transactions on Parallel and Distributed Systems}, vol.~33, no.~11, pp. 2968--2978, Nov. 2022.

\end{thebibliography}
\end{document}